\documentclass[12pt]{article}
\usepackage{amssymb}
 \usepackage{graphicx}
 \usepackage[cp1251]{inputenc}
 \usepackage{rotating}
\begin{document}
 \noindent {\footnotesize\it Astronomy Letters, 2025, Vol. 51, No. 5, p. 278--286}
 \newcommand{\dif}{\textrm{d}}

 \noindent
 \begin{tabular}{llllllllllllllllllllllllllllllllllllllllllllll}
 & & & & & & & & & & & & & & & & & & & & & & & & & & & & & & & & & & & & & &\\\hline\hline
 \end{tabular}

 \vskip 0.2cm
 \bigskip
 
  \centerline{\bf\Large  The Spiral Pattern Speed in the Milky Way Galaxy}
 
 \bigskip
 \centerline{\bf   V. V. Bobylev, A. T. Bajkova, and A. A. Smirnov}
 \bigskip
 \centerline{\it Central (Pulkovo) Astronomical Observatory, Russian Academy of Sciences}
 \bigskip
 \bigskip
{For a sample of masers, the basic kinematic equations were solved by including the Galactic rotation parameters and the peculiar velocity of the Sun as the unknown variables. Based on spectral analysis, the following estimates were obtained: $|f|_{R,\theta}=(7.0,5.1)\pm(1.2,1.4)$~km s$^{-1}$ and the corresponding wavelengths $\lambda_{R,\theta}=(1.9,1.7)\pm(0.4,0.7)$~kpc, as well as $\chi_\odot=-140^\circ\pm15^\circ$. The presence of periodic perturbations in the vertical velocities of masers with an amplitude of $|f|_W=3.1\pm1.4$~km s$^{-1}$ and a wavelength of $\lambda=1.9\pm0.8$~kpc was confirmed. It is shown that the velocities $f_R$ and $f_\theta$ can have both the same and different signs. Therefore, we obtained a large scatter of estimates. Thus, if $f_R$ and $f_\theta$ have the same signs, then $\Omega_p=25.8\pm2.0$~km s$^{-1}$ kpc$^{-1}$ and $R_{cor}=9.1\pm0.8$~kpc. And when $f_R$ and $f_\theta$ have different signs, then $\Omega_p=35.4\pm2.0$~km s$^{-1}$ kpc$^{-1}$ and $R_{cor}=6.8\pm0.8$~kpc.
  }

 \bigskip
 \section*{INTRODUCTION}
The spiral structure in the disk of our Galaxy has been actively studied by various methods for a long time. However, the specific details of this process are currently not well understood (Sellwood 2011; Dobbs, Baba 2014; Bland-Hawthorne, Gerhard 2016; Sellwood, Masters 2022). For example, there is no unambiguous answer to the question of the number of spiral arms $m$, there is a wide range of estimates of the angular velocity of the spiral pattern $\Omega_p$, which makes it difficult to determine the exact position of the corotation radius $R_{\rm cor}$, etc.

The linear density wave theory of Lin, Shu (1964) is quite widely used. In this theory, the wave is described by a logarithmic spiral with a twist angle of $i$ and has $m$ spiral arms. The spiral pattern rotates rigidly with an angular velocity of $\Omega_p$. The density wave disturbances in radial and tangential velocities are periodic with amplitudes of $f_R$ and $f_\theta$, the moduli of which do not exceed approximately 10~km s$^{-1}$. Based on this theory, Yuan (1969) obtained one of the first estimates of $\Omega_p=13.5$~km s$^{-1}$ kpc$^{-1}$.

 Starting with the pioneering works of Lin, Shu (1964), Lin et al. (1969), Yuan (1969), a huge number of scientific publications have been devoted to the problem of applying the linear theory of spiral density waves to the analysis of real data (Marochnik et al. 1972; Cr\'ez\'e, Mennesier 1973; Byl, Ovenden 1978; Mishurov et al. 1979; Loktin, Matkin 1992; Mishurov et al. 1997; Mishurov, Zenina 1997; Amaral, L\'epine 1997; Rastorguev et al. 2001; Melnik et al. 2001; Fernandez et al. 2001; Diaz, L\'epine 2005; Siebert et al. 2012; Griv et al. 2013; Silva, Napiwotsky 2013; Junqueira et al. 2015; Dambis et al. 2015; Diaz et al. 2019; Castro-Guinard et al. 2021; Bobylev, Bajkova 2023a; 2023b).

The exact values of the angular velocity of the spiral pattern and the corotation radius are of great interest for understanding the structure of the galactic disk. However, modern estimates of these parameters lie in a wide range. For example, in the paper by Gerhard (2011), devoted to a review of $\Omega_p$ values, it was concluded that the $\Omega_p$ value is slightly less than the rotation velocity of the Galaxy at the circumsolar distance $R_0$. This means that the corotation radius is slightly further than $R_0$. The average $\Omega_p$ value found from indicators with ages of $10^7-10^8$~years is $25.2$~km s$^{-1}$ kpc$^{-1}$. It should be noted, however, that works devoted to the distribution of stellar velocities in the solar neighborhood give a wider range of $\Omega_p$ values: 17--33~km s$^{-1}$ kpc$^{-1}$. That is, according to some definitions, the value of $\Omega_p$ may exceed the rotation speed of the Galaxy $\Omega_0: 26-30$~km s$^{-1}$ kpc$^{-1}$, then the corotation radius should lie closer to the center of the Galaxy than the Sun.

According to various authors, the number of spiral arms in the Galaxy $m$ varies from 1 to 12. The most commonly proposed values are $m=1, 2, 4$. The single-arm model ($m=1$) of the logarithmic spiral pattern of the Galaxy is substantiated in the works of Griv et al. (2013; 2014; 2017; 2021). The two-arm model ($m=2$) is often used both in numerical modeling and in direct estimates of the values of the sought parameters using observational data (e.g., Byl, Ovenden 1978; Mishurov et al. 1979). The three-arm model ($m=3$) of the spiral structure of the Galaxy was tested, for example, in the works of Russeil (2003), Hou, Han (2014). In this case, the authors usually opt for the four-arm model. In the reviews by Vall\'ee (2008; 2013; 2015; 2017), one can find summaries of the values of the main parameters of the spiral structure of the Milky Way. In particular, arguments are given in favor of the implementation of a four-armed spiral pattern with a twist angle of about $-13^\circ$ in the Galaxy. Reid et al. (2014; 2019) are also inclined to the four-arm model based on the analysis of maser sources with high-precision measurements of their trigonometric parallaxes obtained as a result of VLBI observations. Moreover, these authors propose a four-arm model with a sector structure of the spiral arms. A number of composite models are known. For example, the $m=2+4$ model proposed in the works of Amaral, L\'epine (1997) and L\'epine et al. (2001) in the immediate vicinity of the Sun consists of a superposition of a two-arm and a four-arm patterns. A different model, $m=2+4$, was developed by Xu et al. (2023), where the spiral pattern in the inner region of the Galaxy is two-armed, which then transforms into a four-armed one. The value $m=12$ was proposed by Loktin, Popova (2007) based on an analysis of the distribution of open star clusters, classical Cepheids, and HII zones in the galactic plane.

Currently, there is no clear answer to the question of the exact value of the twist angle of the spiral pattern $i$ in the Galaxy.
In the models of Griv et al. (2013; 2014; 2017; 2021), for $m=1$, the value of $i$ is about $-1^\circ$. In the models of Vall\'ee (1995; 2002; 2008; 2017), Bobylev, Bajkova (2014), for $m=4$, the value of $i$ for all four arms is one, constant and $-13^\circ$. Reid et al. (2014; 2019), based on the analysis of masers with measured VLBI parallaxes and trigonometric parallaxes, propose a model $m=4$, but with a sectoral structure of the spiral arms, where each of the arms consists of sectors with different geometric characteristics. In this model, the largest jump in the arm shape is seen near the Carina--Sagittarius arm in the $l\sim30^\circ$ direction. A similar model was previously constructed by Cordes, Lazio (2002) using pulsar data. It should be noted that in other galaxies, the twist angle is often variable (Savchenko, Reshetnikov 2013).

Thus, when analyzing the same or similar data, a wide variety of values for the geometric parameters of the spiral pattern are obtained. We can agree with the opinion of Xu et al. (2023) that the overall spiral structure of the Milky Way currently remains somewhat uncertain.

In addition to the linear density wave theory of Lin, Shu (1964), a number of models with different physics are known, in which the spiral pattern is non-stationary. Transient spirals are characterized by a short lifetime, 1--2 revolutions around the galactic center (e.g., Sellwood 2011). Such spirals rotate with an angular velocity $\Omega_p (R)$, which depends on the galactocentric distance $R$, the twist angle of the spiral pattern is also not constant (e.g., Sellwood, Masters 2022). Models of short-lived spiral arms were proposed in the works of Goldreich, Lynden-Bell (1965), Julian, Toomre (1966), and Toomre (1981). These studies argue that the spiral structure can be formed due to local instabilities and shear rotation. In this case, the spiral arms are restored due to a local increase in instability.

It is interesting to note that some theoretical models predict a vertical motion of the stellar mass of the disk caused by spiral arms. Asano et al. (2024) analyzed the ``breathing motion'' -- a coherent asymmetric vertical motion around the midplane of the Milky Way disk -- using stellar data from the Gaia~DR3 catalog (Gaia Collaboration 2023). It was found that the compressive ``breathing motion'' manifests itself along the Local Arm (the Orion Arm). According to Asano et al. (2024), the compression of the Local Arm is explained by the fact that it is in the growth phase of a transitive and dynamic spiral arm.

The aim of this paper is to redefine $\Omega_p$ and the location of the corotation radius $R_{\rm cor}$ based on recent data on masers with trigonometric parallaxes measured by the VLBI method. This is achieved by using our original approach, based on spectral analysis of the radial and residual tangential velocities of masers to estimate the amplitudes of the velocity perturbations caused by the influence of the spiral density wave.

\section*{DATA}
Maser emission occurs in the immediate vicinity of young, forming stars, as well as mature stars, fueled either by intense infrared radiation or by violent gas collisions in disks, jet streams, or winds. Masers are characterized by the fact that virtually all of their energy is emitted in a few molecular lines. Examples include hydroxyl (OH) masers with a frequency of 1.6 GHz, methanol (CH$_3$OH) masers with frequencies of 6.7 and 12.2 GHz, water vapor (H$_2$O) masers with a frequency of 22 GHz, and silicon monoxide (SiO) masers with a frequency of 43 GHz.

The use of VLBI to measure the trigonometric parallaxes of galactic masers with relative errors averaging less than 10\% has made them first-class targets for studying the structure and kinematics of the Galaxy. The value of radio observations lies in the fact that they are not hampered by absorption by interstellar dust. Of greatest interest are maser sources associated with young stars and protostars located in regions of active star formation.

Bobylev (2024) reviewed 308 VLBI parallax and proper motion measurements of maser sources and radio stars. A list of characteristics of these objects, supplemented by the latest measurements, is given in Table~\ref{T-mas}, where the first few lines are listed; the remainder will be available online. The stellar radial velocities in this table are given relative to the local standard of rest.

G209.60+01.19 (IRAS 06469+0333). VLBI measurements of this source were carried out within the framework of the Japanese VERA (VLBI Exploration of Radio Astrometry \footnote {http://veraserver.mtk.nao.ac.jp}) program and published in the paper by Loungrueang et al. (2025). Spots of H$_2$O masers at a frequency of 22 GHz were observed in this source from January 2013 to October 2014. VLBI measurements of the trigonometric parallax of this object are carried out for the first time, but they are not yet distinguished by high accuracy, since the relative error in determining the parallax here is about 27\%.

G031.56+05.33 (EC~95). This is a young stellar system consisting of three components EC~95A, EC~95B, and EC~95C with a total mass of $M_{A+B+C}=4.76^{+0.45}_{-0.36} M_\odot$ (Ord\'o\~nez-Toro et al. 2025a). Radio observations of this stellar system are carried out in the continuum at frequencies of 8.4 and 4.9 GHz as part of the GOBELINS (Gould's Belt Distances Survey, Ortiz-Le\'on et al. 2018) and DYNAMO–VLBA (Dynamical Masses of Young Stellar Multiple Systems with the VLBA) programs. RSBD measurements of the trigonometric parallax of this system were known previously (e.g., Ortiz-Le\'on et al. 2018). In Ord\'o\~nez-Toro et al. (2025)a, the parallax and proper motion components of this system were improved by using a twelve-year series of RSBD observations conducted from 2007 to 2019.

G353.10+16.89 (Rho Oph S1). This close binary system, consisting of components A and B with masses of $4.115\pm0.039 M_\odot$ and $0.814\pm0.006 M_\odot$, respectively, also has a long history of VLBI observations, conducted in the continuum at frequencies of 8.4 and 4.9 GHz. The trigonometric parallax and motion of the center of mass of this system were refined in Ord\'o\~nez-Toro et al. (2025)b. The dynamical masses were estimated by constructing the orbits of the components about their common center of mass.

G008.83-00.02, G010.47+00.02 and G024.78+00.08. The first VLBA measurements of these three sources were made as part of the BeSSeL (The Bar and Spiral Structure Legacy Survey \footnote {http://bessel.vlbi-astrometry.org}) program using the American VLBA array. The measurement results were published in the paper by Kumar et al. (2025). The accuracy of determining the paprallaxes here is very high ($\sigma_\pi\ / \pi<7\%$). All three of these sources are located quite far from the Sun (further than 4.8 kpc) and are associated with the so-called 3-kpc spiral arm.

   \begin{table}[t]
   \caption[]{\small  Data on maser sources and radio stars }
   \begin{center}  \label{T-mas}    \small
       \begin{tabular}{|l|r|r|r|r|r|r|r|r|r|r|r|}      \hline
Source & $\alpha$~~~ & $\delta$~~~ & $\pi(\sigma_\pi)$ &  $\mu^*_\alpha (\sigma_{\mu_\alpha})$ & $\mu_\delta(\sigma_{\mu_\delta})$ & $V_r (\sigma_{V})$ & Ref \\
  & deg & deg & mas &  mas yr$^{-1}$ & mas yr$^{-1}$ & km s$^{-1}$ &   \\\hline
G209.60+01.19  &102.4050 & $   3.5084$ & $0.618 (.166)$ & $ -1.15 (.07)$ & $-0.09 (.61)$ & $30 (5)$ & (1) \\
G031.56+05.33  &277.4912 & $   1.2128$ & $2.30  (.04)$   & $  3.54 (.02)$ & $-8.42 (.02)$ & $  9  (3)$ & (2) \\
G353.10+16.89  &246.6424 & $-24.3912$ & $7.30  (.02)$   & $ 2.486 (.002)$ & $-26.756 (.001)$ & $  3  (3)$ & (3) \\
G008.83$-$00.02  &271.3570 & $-21.3237$ & $0.208 (.019)$  & $-1.35 (.27)$ & $-1.43 (.46)$ & $   1 (5)$ & (4) \\
G010.47+00.02 &272.1593 & $-19.8640$ & $0.115 (.008)$  & $-3.69 (.08)$ & $-6.34 (.07)$ & $ 69  (5)$ & (4) \\
G024.78+00.08 &279.0523 & $-07.2030$ & $0.188 (.007)$  & $-2.60 (.12)$ & $-4.55 (.30)$ & $110 (5)$ & (4) \\
     & \multicolumn{5}{c}{...} & &  \\\hline
     \end{tabular} \end{center}
     {\small
   (1)~Loungrueang et al. (2025);
   (2)~Ord\'o\~nez-Toro et al. (2025)a;
   (3)~Ord\'o\~nez-Toro et al. (2025)b;
   (4)~Kumar et al.. (2025).
    }
   \end{table}

 \section{METHODS}\label{method}
Our approach is based on solving a system of equations of the following type:
 \begin{equation}
 \begin{array}{lll}
 V_r=-U_\odot\cos b\cos l-V_\odot\cos b\sin l -W_\odot\sin b\\
   +R_0(R-R_0)\sin l\cos b\Omega^\prime_0
  +0.5R_0(R-R_0)^2\sin l\cos b\Omega^{\prime\prime}_0\\
   -f_R\cos\chi\cos(l+\theta)\cos b+f_\theta\sin\chi\sin(l+\theta)\cos b,
 \label{EQ-1}
 \end{array}
 \end{equation}
 \begin{equation}
 \begin{array}{lll}
 V_l= U_\odot\sin l-V_\odot\cos l
  -r\Omega_0\cos b\\
  +(R-R_0)(R_0\cos l-r\cos b)\Omega^\prime_0
  +0.5(R-R_0)^2(R_0\cos l-r\cos b)\Omega^{\prime\prime}_0\\
  +f_R\cos\chi\sin(l+\theta)+f_\theta\sin\chi\cos(l+\theta),
 \label{EQ-2}
 \end{array}
 \end{equation}
 \begin{equation}
 \begin{array}{lll}
 V_b=U_\odot\cos l\sin b + V_\odot\sin l \sin b-W_\odot\cos b\\
  -R_0(R-R_0)\sin l\sin b\Omega^\prime_0
  -0.5R_0(R-R_0)^2\sin l\sin b\Omega^{\prime\prime}_0\\
  +f_R\cos\chi\cos(l+\theta)\sin b -f_\theta\sin\chi\sin(l+\theta)\sin b,
 \label{EQ-3}
 \end{array}
 \end{equation}
where $V_r$ is the radial velocity of the star, $V_l=4.74r\mu_l\cos b$ and $V_b=4.74r\mu_b$ are the projections of its tangential velocity, which are calculated using the heliocentric distance to the star $r=1\pi$, calculated through the trigonometric parallax $\pi$, and the corresponding values of the components of proper motion $\mu_l$ and $\mu_b$, $R$ ~--- the distance of the star from the rotation axis of the Galaxy $R^2=r^2\cos^2 b-2R_0 r\cos b\cos l+R^2_0,$ the velocities $(U,V,W)_\odot$ are the average group velocity of the sample and reflect the peculiar motion of the Sun relative to the stars used, $\Omega_0$ is the angular velocity of rotation of the Galaxy at the solar distance $R_0$, $\Omega^{\prime}_0$ and $\Omega^{\prime\prime}_0$ are the corresponding derivatives of the angular velocity, $f_R$ and $f_\theta$ are the amplitudes of the perturbations of the radial and tangential velocities, respectively. The value of $R_0$ is taken to be $8.1\pm0.1$~kpc according to the review by Bobylev, Bajkova (2021).

\subsection{First Method}
According to the linear density wave theory of Lin, Shu (1964), perturbations from the galactic spiral density wave
in the velocities $V_R$ and $\Delta V_{circ}$ are periodic and are described by functions of the following type:
\begin{equation}
\begin{array}{lll}
V_R=f_R\cos\chi,\quad \Delta V_{\rm circ}=f_\theta\sin\chi,
\label{DelVRot}
\end{array}
\end{equation}
where $\Delta V_{circ}$ is the tangential velocity of the star, independent of galactic rotation,
$\chi=m[\cot(i)\ln(R/R_0)-\theta]+\chi_\odot$ is the radial phase of the spiral wave, and $\chi_\odot$ is the radial phase of the Sun in the spiral wave, $m$ is the number of spiral arms, $i$ is the twist angle of the spiral pattern ($i<0$ for twisting spirals). The wavelength $\lambda$ is calculated based on the ratio $2\pi R_0/\lambda=m\cot(|i|)$.

The first method is that $f_R$ and $f_\theta$ are found directly from the least-squares (LS) solution of a system of conditional equations of the form~(\ref{EQ-1})--(\ref{EQ-3}) along with the remaining determined parameters: $U_\odot,$ $V_\odot,$ $W_\odot,$ $\Omega_0,$ $\Omega^\prime_0,$ $\Omega^{\prime\prime}_0.$ The solution is sought with weights of the form $w_{r,l,b}=S_0/\sqrt {S_0^2+\sigma^2_{V_{r,l,b} }},$ where $S_0$~--- ``cosmic'' variance, $\sigma_{V_r}, \sigma_{V_l}, \sigma_{V_b}$~--- the variance of the errors in the observed velocities. The value of $S_0$ for masers is taken to be $10$~km s$^{-1}$. The solution is performed in several iterations using the $3\sigma$ criterion to exclude equations with large residuals.

The angular velocity of rotation of the spiral pattern $\Omega_p$ is estimated as follows.
According to Lin et al. (1969), $f_R$ and $f_\theta$ have the following form:
 \begin{equation}
 \renewcommand{\arraystretch}{2.0}
      f_R= {k A\over {\varkappa} }  {\nu \over {1-\nu^2}} \Im^{(1)}_\nu(x),
 \label{Factors-1}
 \end{equation}
 \begin{equation}
   f_\theta= -{k A\over 2\Omega} {1 \over {1-\nu^2}} \Im^{(2)}_\nu(x),
 \label{Factors-2}
 \end{equation}
where
$A$ is the amplitude of the spiral wave potential, which we estimate based on the well-known relation (Fernandez et al. 2008)
$A=(R_0\Omega_0)^2 f_{r0} \tan (|i|)/m$ with the value $f_{r0}=0.04\pm0.01$ (the ratio of the radial component of the gravitational force corresponding to the spiral arms to the total gravitational force of the Galaxy, $k=m \cot i/R$ is the radial wave number, $\Omega=\Omega(R)$ is the angular velocity of the Galaxy's rotation,
$\varkappa^2=4\Omega^2\left(1+\frac{\displaystyle R}{\displaystyle 2\Omega}\frac{\displaystyle d\Omega}{\displaystyle dR}\right)$~--- epicyclic frequency ($\varkappa>0$),
$\nu=m(\Omega_p-\Omega)/\varkappa$~--- the frequency with which a test particle encounters a passing spiral perturbation,
$\Im^{(1)}_\nu(x)$ and $\Im^{(2)}_\nu(x)$~--- reduction factors which are functions of the coordinate $x=k^2\sigma^2_R/\varkappa^2,$ where $\sigma_R$~--- the root-mean-square dispersion of stellar radial velocities.

If the ratio of the reduction factors $\Im^{(2)}_\nu/\Im^{(1)}_\nu$ is known, then $\Omega_p$ for $R=R_0$ and $\Omega=\Omega_0$ can be can be estimated as follows:
\begin{equation}
\renewcommand{\arraystretch}{2.0}
\Omega_p-\Omega_0=-\frac{\Im^{(2)}_\nu}{\Im^{(1)}_\nu}~
\frac{f_R}{f_\theta}~ \frac{2\Omega_0}{m}\biggr(1+\frac{R_0\Omega'_0}{2\Omega_0}\biggr).
\label{Im1-Im2}
\end{equation}
According to the calculations of Cr\'ez\'e, Mennesier (1973), $\Im^{(1)}_\nu>0$ for any values of $x=k^2\sigma^2_R/\varkappa^2,$ and $\Im^{(2)}_\nu>0$ for small $x$, but can take negative values for large $x$. Mishurov et al. (1997), when analyzing a sample of long-period (i.e., young) Cepheids, found that $\Im^{(1)}_\nu>0$ and $\Im^{(2)}_\nu>0$. In the paper by Bobylev, Bajkova (2023a), a positive value of $\Im^{(2)}_\nu$ was also found for a sample of masers, and the ratio of the reduction factors $\Im^{(2)}_\nu/\Im^{(1)}_\nu=0.672$ was calculated.

\subsection{Second Method}
The second method involves estimating only six unknowns from the least-squares solution of the system of conditional equations~(\ref{EQ-1})--(\ref{EQ-3}): $U_\odot,$ $V_\odot,$ $W_\odot,$ $\Omega_0,$ $\Omega^\prime_0$, and $\Omega^{\prime\prime}_0$. Using the determined parameters of the Galaxy's rotation, we calculate the residual velocities $\Delta V_{\rm circ}$. And, based on spectral analysis, we determine the values of $i,$ $\chi_\odot,$ $f_R$, and $f_\theta$. The spectral analysis we use is described in detail in Bajkova, Bobylev (2012). Its main feature is that it takes into account the positional angles of objects and the logarithmic nature of the spiral wave.

Note that spectral analysis only allows us to obtain information about the velocity moduli $|f|_{R,\theta, W}$. Following the theory of Lin, Shu (1964), and taking into account the explanations of Rolfs (1977), we have the following working relationships:
\begin{equation}
 \begin{array}{lll}
                 V_R=-|f_R|\cos\chi,\quad  \Delta V_{\rm circ}=-|f_\theta|\sin\chi,
 \label{DelVRot}
 \end{array}
 \end{equation}
where the minus sign of $|f_R|$ means that at the center of the spiral arm (e.g., for $\chi=0^\circ$) this velocity is directed toward the galactic center, while the minus sign of $|f_\theta|$ means that at the inner edge of the spiral arm (e.g., for $\chi=\pi/2$) this velocity is directed counter to the galactic rotation.

The theory of Lin, Shu (1964) considers an infinitely thin disk, where there are perturbations of only two types of velocities, $f_R$ and $f_\theta$. However, a number of modern studies (Bobylev, Bajkova 2015; Rastorguev et al. 2017; Loktin, Popova 2019) have shown that there are also small-amplitude perturbations of the vertical velocities $W$ with a wavelength characteristic of the galactic density wave $\lambda\sim2$~kpc. Therefore, we also included vertical velocities in our analysis.

\begin{figure}[t]
{ \begin{center}
  \includegraphics[width=0.88\textwidth]{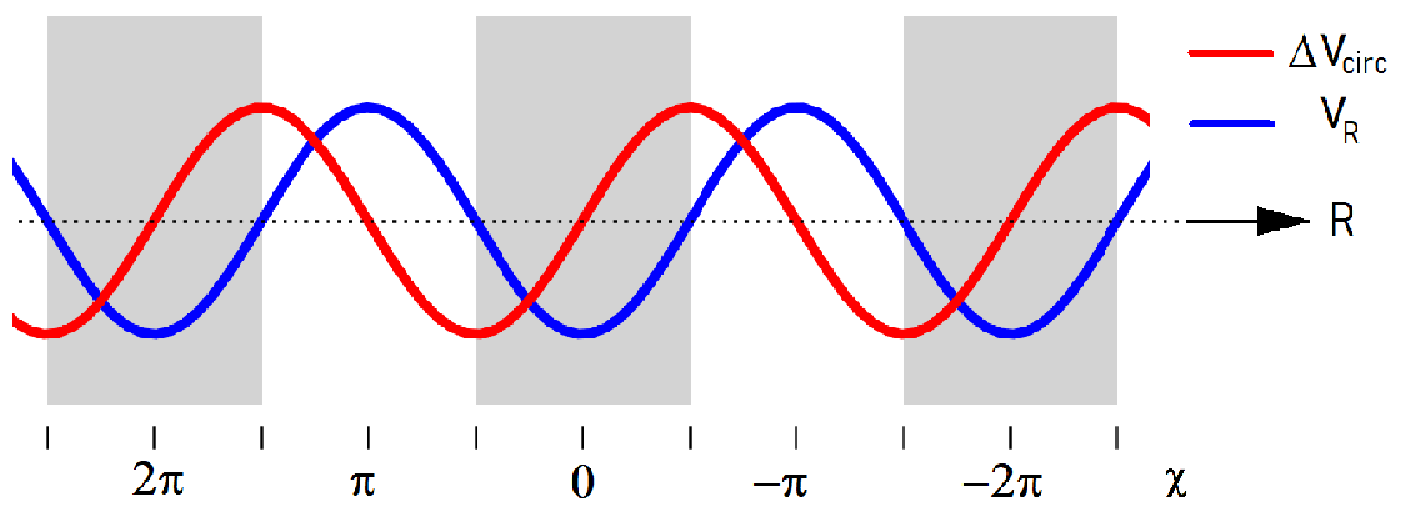}
  \caption{Dependences of the velocities $V_R$ and $\Delta V_{circ}$ on $R$ according to the linear theory of density waves of Lin, Shu (1964), constructed using the data of Fig. 21 from the monograph by Rolfs (1977) for the case $\Omega_p<\Omega(R)$, where the $R$ axis is directed from the center of the Galaxy towards the Sun, and the radial phase $\chi$ is directed toward the center of the Galaxy.
  }
 \label{f-sin-cos}
\end{center}}
\end{figure}

We have prepared Fig.~\ref{f-sin-cos} with the dependences of the velocities $V_R$ and $\Delta V_{circ}$ on the distance $R$ for the case where these velocities are consistent with the provisions of the linear density wave theory of Lin, Shu (1964). This figure is constructed based on the data in Fig.~21 from the monograph by Rohlfs (1977). Using the dependences of the velocities $V_R$ and $\Delta V_{circ}$ on $R$, constructed from observational data, we can determine the positions of the centers of the spiral arms and the position of the Sun in the wave, i.e., estimate the value of the solar phase $\chi_\odot$.

The relations~(\ref{DelVRot}) when using the results of spectral analysis are very important for correctly estimating the value of the angular velocity $\Omega_p$. In particular, from the expression (\ref{Im1-Im2}) one can see that the sign of the correction $\Delta\Omega$ to $\Omega_0$ depends on the signs of the velocities $f_R$ and $f_\theta$.

Note that residual velocities can be either ``initial'' or acquired according to the transient spiral arm model (e.g., Grand, Kawata 2016). However, in our models, we do not take into account the gravitational influence of the spiral arm on the stars, as we work within the framework of density wave theory.

Given an estimate of $\Omega_p$, we find the corotation radius $R_{\rm cor}$. This value is calculated from the relationship obtained by equating the linear rotational velocities of the Galaxy and the spiral pattern:
$R_{\rm cor}=R_0+(\Omega_p-\Omega_0)/\Omega'_0.$

\begin{figure}[t]
{ \begin{center}
  \includegraphics[width=0.6\textwidth]{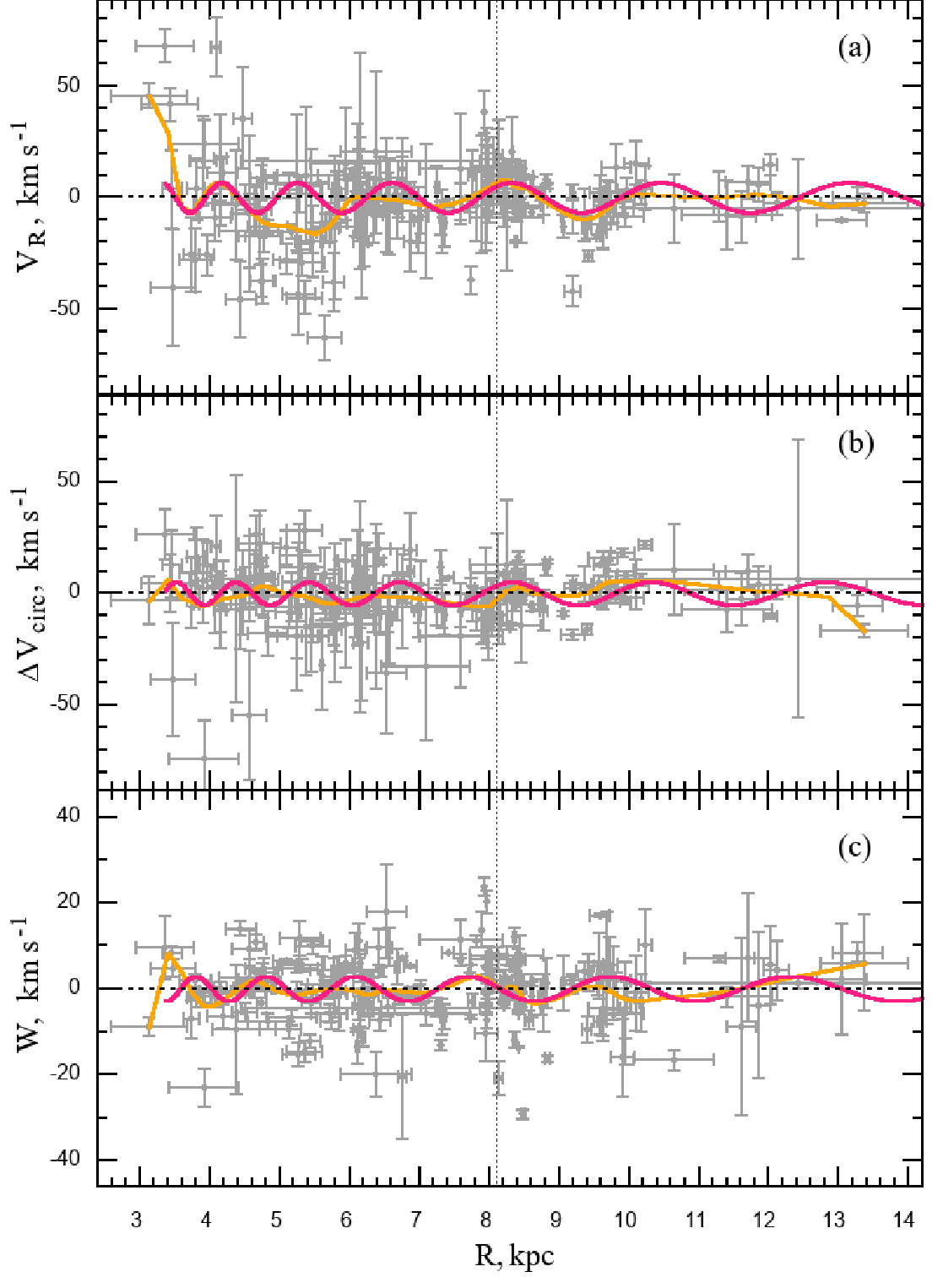}
  \caption{Radial $V_R$ (a), residual tangential $\Delta V_{circ}$ (b) and vertical $W$ (c) maser velocities as a function of distance $R$, red lines are periodic curves found based on spectral analysis, orange lines are smoothed averages.}
 \label{f-V-rtw}
\end{center}}
\end{figure}
\begin{figure}[t]
{ \begin{center}
  \includegraphics[width=0.96\textwidth]{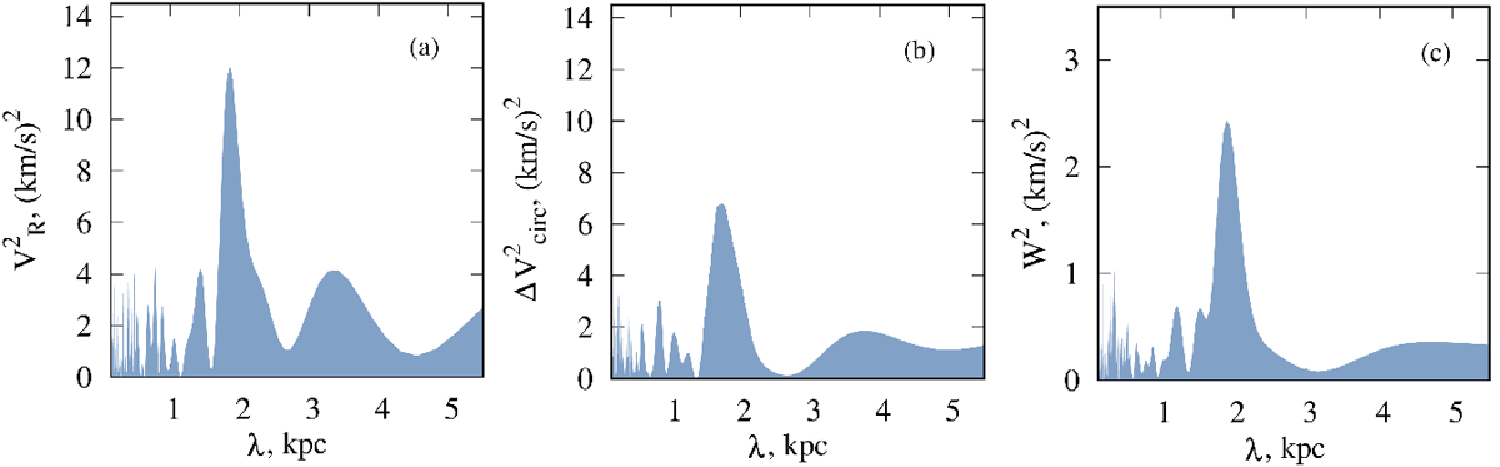}
  \caption{Power spectra of radial (a), residual tangential (b) and vertical (c) velocities of masers.}
 \label{f-Sp}
\end{center}}
\end{figure}

\section{RESULTS AND DISCUSSION}\label{result}
We used 212 masers with trigonometric parallax measurement errors of less than 15\%, located in the region $3<R<14$~kpc. Using this approach, we found velocities $(U,V,W)_\odot=(8.42,13.19,8.49)\pm(0.86,0.90,0.70)$~km s$^{-1}$, as well as:
\begin{equation}  \label{sol-212}
  \begin{array}{lll}
      \Omega_0 =~29.66\pm0.28~\hbox{km s$^{-1}$ kpc$^{-1}$},\\
  \Omega^{'}_0 =-4.268\pm0.063~\hbox{km s$^{-1}$ kpc$^{-2}$},\\
 \Omega^{''}_0 =~0.805\pm0.033~\hbox{km s$^{-1}$ kpc$^{-3}$},
 \end{array} \end{equation}
where $\sigma_0=10.4$~km s$^{-1}$, the linear rotation velocity of the local standard of rest was
$V_0=240.2\pm3.8$~km s$^{-1}$ (for the adopted $R_0=8.1\pm0.1$~kpc).

We note the good agreement between the parameter values found in solution~(\ref{sol-212}) and the results of maser analysis by other authors. Thus, based on a sample of 131 masers, the following values were obtained by Rastorguev et al. (2017) (model A1): $\Omega_0=28.94\pm0.51$~km s$^{-1}$ kpc$^{-1}$, $\Omega^{'}_0 =-3.91\pm0.07$~km s$^{-1}$ kpc$^{-2}$ and
$\Omega^{''}_0 =~0.86\pm0.03$~km s$^{-1}$ kpc$^{-3}$, as well as $R_0=8.21\pm0.12$~kpc.
In the paper by Reid et al. (2019), the angular velocity was $\Omega_0=29.0\pm0.3$~km s$^{-1}$ kpc$^{-1}$, and $R_0=8.15\pm0.15$~kpc. Based on a large sample of 201 maser sources, in a recent paper by Nikiforov (2025), $(U,V,W)_\odot=(6.09,15.55,8.22)\pm(0.86,0.92,0.92)$~km s$^{-1}$, $\Omega_0=28.43\pm0.22$~km s$^{-1}$ kpc$^{-1}$, and $R_0=7.88\pm0.12$~kpc.

Fig. ~\ref{f-V-rtw} shows the radial, residual tangential, and vertical velocities of masers as functions of distance $R$.
Plotted are the periodic curves found from spectral analysis, as well as the smoothed averages. We see good agreement between the two types of curves in the range of approximately $7<R<10$~kpc. The best agreement is demonstrated by the $V_R$ velocities.

Figure~\ref{f-Sp} shows the power spectra of the corresponding velocities. Based on the data used to construct this figure, the amplitudes (the moduli of these velocities, kpc, were already noted above) of the disturbance velocities were found: $|f|_{R,\theta,W}=(7.0,5.1,3.1)\pm(1.2,1.4,1.4)$~km s$^{-1}$ and the corresponding wavelengths $\lambda_{R,\theta,W}=(1.9,1.7,1.9)\pm(0.4,0.7,0.8)$~kpc. The errors in determining these quantities were estimated using the Monte Carlo method.

A comparison of the corresponding periodic curves shown in Figure~\ref{f-sin-cos} and Figure~\ref{f-V-rtw} reveals, at first glance, satisfactory agreement between them. Thus, in Fig.~\ref{f-V-rtw}(a) the minimum of the velocity $V_R$ at $R\sim7$~kpc is clearly visible, so we estimated the value of the solar phase $\chi_\odot=-140^\circ\pm15^\circ$. To estimate the correction $\Delta\Omega=\Omega_p-\Omega_0$ (see relation (\ref{Im1-Im2})) the signs of the velocities $f_R$ and $f_\theta$ are important. On the solar circle ($R=R_0$) we find $V_R= -7\cos(-140^\circ)=5.4$~km s$^{-1}$ and $\Delta V_{\rm circ}=-5.1\sin(-140^\circ)=3.3$~km s$^{-1}$. That is, we have both positive quantities in agreement with Fig.~\ref{f-sin-cos} for a given value of the phase $\chi_\odot$. Nevertheless, perfect agreement with the prediction of the linear theory of the spiral density wave of Lin, Shu (1964) for the case shown in Fig.~\ref{f-sin-cos} (as well as Fig.~21 from the monograph by Rohlfs (1977) for $\Omega_p<\Omega(R)$) in our case is achieved only for radial velocities $V_r$. The velocity wave $\Delta V_{\rm circ}$ has a noticeable difference (in phase shift) from the model prediction (Fig.~\ref{f-sin-cos}). Note that satisfactory agreement with Fig.~\ref{f-sin-cos} was found by us in the work of Bobylev, Bajkova (2013) during the analysis of 73 masers. However, expanding the sample of masers revealed increasingly large differences in the phases of the tangential velocities.

Using the $f_R $ and $f_\theta$ perturbation velocities found from spectral analysis (if they have the same signs), we obtained the following new estimates: $\Omega_p=23.96\pm1.98$~km s$^{-1}$ kpc$^{-1}$ and the corotation radius $R_{cor}=9.4\pm0.8$~kpc.

Next, using 212 masers in the range $3<R<14$~kpc, we obtained a solution to kinematic equations of the form (\ref{EQ-1})--(\ref{EQ-3}), assuming $m=4, i=-10^\circ$, $\lambda=2.2$~kpc, and $\chi_\odot=-140^\circ$. Using this approach, we found the velocities $(U,V,W)_\odot=(7.12,13.00,8.84)\pm(0.79,0.73,0.58)$~km s$^{-1}$, and also:
\begin{equation} \label{sol-212-kinematic}
\begin{array}{lll}
\Omega_0 =~29.87\pm0.32~\hbox{km s$^{-1}$ kpc$^{-1}$},\\
\Omega^{'}_0 =-4.458\pm0.062~\hbox{km s$^{-1}$ kpc$^{-2}$},\\
\Omega^{''}_0 =~0.884\pm0.029~\hbox{km s$^{-1}$ kpc$^{-3}$},\\
f_R =-6.75\pm0.94~\hbox{km s$^{-1}$},\\
f_\theta =~0.87\pm0.98~\hbox{km s$^{-1}$},
\end{array} \end{equation}
where $\sigma_0$ has decreased compared to the previous solution~(\ref{sol-212}) to 8.2~km s$^{-1}$, indicating the real benefit of taking into account the influence of the spiral density wave. The linear rotational velocity of the local standard of rest in this solution was $V_0=242.0\pm3.9$~km s$^{-1}$.

The values of $f_R$ and $f_\theta$ found in the solution~(\ref{sol-212-kinematic}) differ little from
$f_R=-6.8\pm1.4$~km s$^{-1}$, $f_\theta=3.1\pm1.0$~km s$^{-1}$, and $\chi_\odot=-123^\circ\pm10^\circ$ obtained in the work of Rastorguev et al. (2017) for model A1, where $f_R$ and $f_\theta$ appeared as the unknowns sought
in solving kinematic equations of the form (\ref{EQ-1})--(\ref{EQ-3}).

When the perturbation velocity values we found, $|f|_{R,\theta}=(7.0,5.1)$~km s$^{-1}$, have different signs, we obtain: $\Omega_p=35.4\pm2.0$~km s$^{-1}$ kpc$^{-1}$ and $R_{cor}=6.8\pm0.8$~kpc.

The status of the Local Arm is important. Some authors (e.g., Loktin, Popova 2019) believe that the Local Arm is part of the Galaxy's global spiral structure (grand design structure). It is well known that the Sun is located at the edge of the Local Arm, so if it is part of the global spiral structure, then the wavelength $\lambda$ in the solar neighborhood should be around 1~kpc. As can be seen from Fig.~\ref{f-Sp}, our results do not support this value of $\lambda$, since the Sun is close to the positive hump of radial velocities, which, according to density wave theory, is characteristic of the inter-arm space.

We found a difference between the calculated phases and Fig.~\ref{f-sin-cos}. Therefore, to correctly determine the signs of $f_R$ and $f_\theta$, it is necessary to know all the necessary conditions. For example, the direction of the density wave. In this regard, we note interesting data recently obtained from an analysis of the distribution of young open star clusters associated with the Radcliffe wave (Konietzka et al. 2024; Bobylev et al. 2025). Specifically, it turned out that the chain of the youngest clusters is located farther from the Sun toward the galactic anticenter than relatively older clusters. In fact, an age gradient has been detected in the Local Arm. And if this age gradient is associated precisely with the spiral density wave, then in this case the rotation radius

\section*{CONCLUSIONS}
A kinematic analysis was conducted on a sample of 212 masers and radio stars for which the trigonometric parallax measurement errors using the VLBI method are less than 15\%. They are located in the region of $3<R<14$~kpc. Application of various analysis methods showed that the galactic rotation parameters are well estimated, while estimates of $\Omega_p$ and the corotation radius $R_{cor}$ lie within a very wide range.

From the solution of the kinematic equations, new estimates of the group velocity of the sample were obtained: $(U,V,W)_\odot=(8.42,13.19,8.49)\pm(0.86,0.90,0.70)$~km s$^{-1}$, as well as the parameters of the angular velocity of the Galaxy's rotation:
$ \Omega_0 =~29.66\pm0.28$~km s$^{-1}$ kpc$^{-1}$,
$\Omega^{'}_0 =-4.268\pm0.063$~km s$^{-1}$ kpc$^{-2}$ and
$\Omega^{''}_0 =~0.805\pm0.033$~km s$^{-1}$ kpc$^{-3}$.
The linear rotation velocity of the local standard of rest relative to the galactic center at a distance around the Sun was $V_0=240.2\pm3.8$~km s$^{-1}$ for the adopted value $R_0=8.1\pm0.1$~kpc.

Further, based on spectral analysis, the following estimates were obtained: $|f|_{R,\theta}=(7.0,5.1)\pm(1.2,1.4)$~km s$^{-1}$ and the corresponding wavelengths $\lambda_{R,\theta}=(1.9,1.7)\pm(0.4,0.7)$~kpc. From these data, $\chi_\odot=-140^\circ\pm15^\circ$. The paradoxical nature of the situation is that if $f_R $ and $f_\theta$ have the same sign, we find $\Omega_p=25.8\pm2.0$~km s$^{-1}$ kpc$^{-1}$ and $R_{cor}=9.1\pm0.8$~kpc. When $f_R$ and $f_\theta$ have different signs, then $\Omega_p=35.4\pm2.0$~km s$^{-1}$ kpc$^{-1}$ and $R_{cor}=6.8\pm0.8$~kpc.
It would be interesting to test the connection between the age gradient of open clusters of stars found in the Radcliffe wave and the spiral wave. Such a gradient may indicate that $\Omega_p>\Omega_0$, and $R_{cor}<R_0$.

The presence of periodic perturbations in the vertical velocities of masers with an amplitude of $|f|_W=3.1\pm1.4$~km s$^{-1}$ and a wavelength of $\lambda=1.9\pm0.8$~kpc was confirmed. This is important, as the linear model of Lin and Shu does not predict vertical velocity perturbations. However, this is consistent with the breathing motion model, which predicts vertical oscillations of stars.

\medskip
The authors are grateful to the reviewer for helpful comments that contributed to improving the article.

\bigskip\medskip{REFERENCES}\medskip {\small
\begin{enumerate}

 \item
L.H. Amaral,  J.R.D. L\'epine, Mon. Not. R. Astron. Soc. {\bf 286}, 885 (1997).

\item
T. Asano, D. Kawata, M.S. Fujii, and J. Baba, Mon. Not. R. Astron. Soc. {\bf 529}, 7 (2024).

\item
A.T. Bajkova, V.V. Bobylev, Astron. Lett. {\bf 38}, 549 (2012).

\item
J. Byl, M.W. Ovenden, Astrophys. J. {\bf 225}, 496 (1978).

\item
J. Bland-Hawthorn, O. Gerhard, Annu. Rev. Astron. Astrophys.  {\bf 54} 529 (2016).

\item
V.V. Bobylev, A.T. Bajkova, Astron. Lett. {\bf 39}, 809 (2013).

\item
V.V. Bobylev, A.T. Bajkova, Mon. Not. R. Astron. Soc. {\bf 437}, 1549 (2014).

\item
V.V. Bobylev, A.T. Bajkova, Mon. Not. R. Astron. Soc. {\bf 447},  L50 (2015).

\item
V.V. Bobylev, A.T. Bajkova, Astron. Rep. {\bf 65}, 498 (2021).

\item
Bobylev V.V., Bajkova A.T.,  Astron. Lett., V. 49, 110 (2023a).

\item
Bobylev V.V., Bajkova A.T.,  Astron. Lett., V. 49, 320 (2023b).

\item
Bobylev V.V.,  Astron. Lett., V. 50, 795 (2024).

\item
V.V. Bobylev, N.R. Ikhsanov, A.T. Bajkova, Astrophys. Bull. {\bf 80}, 181 ( 2025).

\item
A. Castro-Ginard, P.J. McMillan, X. Luri, et al., Astron. Astrophys. {\bf 652}, 162 (2021).

\item
J.M. Cordes, T.J.W. Lazio,  arXiv: astro-ph/0207156 (2002).

\item
M. Cr\'ez\'e, M.O. Mennessier, Astron. Astrophys. {\bf 27}, 281 (1973).

\item
A.K. Dambis, Berdnikov L.N., Efremov Yu.N., et al., Astron. Lett. {\bf 41}, 489 (2015).

\item
W.S. Dias, J.R.D. L\'epine, Astrophys. J.  {\bf 629}, 825 (2005).

\item
W.S. Dias, H. Monteiro, J.R.D. L\'epine, and D.A. Barros, Mon. Not. R. Astron. Soc. {\bf 486}, 5726 (2019).

\item
C. Dobbs, J. Baba, PASA {\bf 31}, 35 (2014).

\item
D. Fern\'andez, F. Figueras, and J. Torra,  Astron. Astrophys. {\bf 372}, 833 (2001).

\item
D. Fern\'andez, F. Figueras, and J. Torra,  Astron. Astrophys. {\bf 480}, 735 (2008).

\item
Gaia Collab. (A. Vallenari, A.G.A. Brown, T. Prusti, et al.), Astron. Astrophys. {\bf 674}, A1 (2023).

\item
O. Gerhard, Mem. S. A. It. Suppl. {\bf 18}, 185 (2011).

\item
P. Goldreich, D. Lynden-Bell, Mon. Not. R. Astron. Soc. {\bf 130}, 125 (1965).

\item
R.J.J. Grand,  D. Kawata, AN {\bf 337}, 957 (2016).

\item
E. Griv, C.-C. Ngeow and I.-G. Jiang, Mon. Not. R. Astron. Soc. {\bf 433}, 2511 (2013).

\item
E. Griv, C.-C. Lin, C.-C. Ngeow and I.-G. Jiang, New Astron. {\bf 29}, 9 (2014).

\item
E. Griv, L.-G. Hou, I.-G. Jiang, and C.-C. Ngeow, Mon. Not. R. Astron. Soc. {\bf 464}, 4495 (2017).

\item
E. Griv, L.-G. Hou, I.-G. Jiang, and C.-C. Ngeow, Mon. Not. R. Astron. Soc. {\bf 464}, 4495 (2021).

\item
L.G. Hou, J.L. Han, Astron. Astrophys. {\bf 569}, 125  (2014).

\item
W. H. Julian, A. Toomre, Astrophys. J.  {\bf 146}, 810 (1966).

\item
T.C. Junqueira, C. Chiappini, J.R.D. L\'epine, et al., Mon. Not. R. Astron. Soc. {\bf 449}, 2336  (2015).

\item
R. Konietzka, A.A. Goodman, C. Zucker, et al.,  Nature {\bf 628}, 62 (2024).

\item
J. Kumar, M.J. Reid, T. M. Dame, et al., Astrophys. J. {\bf 982}, 185 (2025).

\item
C.C. Lin, F.H. Shu, Astrophys. J. {\bf 140}, 646 (1964).

\item
C.C. Lin, C. Yuan, and F.H. Shu, Astrophys. J. {\bf 155}, 721 (1969).

\item
J.R.D. L\'epine,  Yu.N. Mishurov, and S.Yu, Dedikov,  Astrophys. J. {\bf 546}, 234  (2001).

\item
A.V. Loktin, N.V. Matkin, Astron. Astrophys. Tr. {\bf 3}, 169 (1992).

\item
A.V. Loktin, M.E. Popova, Astron. Rep. {\bf 51}, 364 (2007).

\item
A.V. Loktin, M.E. Popova, Astrophys. Bull. {\bf 74}, 270 (2019).

\item
Y. Loungrueang, N. Sakai, K. Sugiyama, and S. Wannawichian,  J. Phys.: Conf. Ser. {\bf 2934}, 012006 (2025).

\item
L.S. Marochnik, Yu.N. Mishurov, and A.A. Suchkov, Astrophys. Space Science {\bf 19}, 285 (1972).

\item
A.M. Melnik, A.K. Dambis, and A.S. Rastorguev, Astron. Lett. {\bf 27}, 521 (2001).

\item
Yu.N. Mishurov, E.D. Pavlovskaia, and A.A. Suchkov, Astron. Rep. {\bf 23}, 147 (1979).

\item
Yu.N. Mishurov, I.A. Zenina, A.K. Dambis, et al., Astron. Astrophys. {\bf 323}, 775 (1997).

\item
Yu.N. Mishurov, I.A. Zenina, Astron. Astrophys. {\bf 341}, 81 (1999).

\item
I. I. Nikiforov,  Solar System Res. {\bf 59}, 35 (2025).

\item
J. Ord\'o\~nez-Toro, S.A. Dzib, L. Loinard, et al., Mon. Not. R. Astron. Soc. {\bf 540}, 2830 (2025)a.

\item
J. Ord\'o\~nez-Toro, S.A. Dzib, L. Loinard, et al., Mon. Not. R. Astron. Soc. {\bf 538}, 1784 (2025)b.

\item
G.N. Ortiz-Le\'on, L. Loinard, S.A. Dzib,  et al.,   Astrophys. J. {\bf 865}, 73 (2018).

\item
A.S. Rastorguev, E.V. Glushkova, M.V. Zabolotskikh, and H. Baumgardt, Astron. Astrophys. Tr. {\bf 20}, 103 (2001).

 \item
A.S. Rastorguev, M.V. Zabolotskikh, A.K. Dambis, et al., Astrophys. Bulletin {\bf 72}, 122 (2017).

\item
M.J. Reid, K.M. Menten, A. Brunthaler, et al.,  Astrophys. J. {\bf 783}, 130 (2014).

\item
M.J. Reid, N. Dame, K.M. Menten, et al., Astrophys. J. {\bf 885}, 131 (2019).

\item
K. Rohlfs, {\it Lectures on Density Wave Theory} (Springer-Verlag, Berlin, 1977).

\item
D. Russeil, Astron. Astrophys. {\bf 397}, 133  (2003).

\item
S.S. Savchenko, V. P. Reshetnikov, Mon. Not. R. Astron. Soc. {\bf 436}, 1074 (2013).

 \item
J. A. Sellwood, Mon. Not. R. Astron. Soc. {\bf 410}, 1637 (2011).

\item
J. A. Sellwood, K. L. Masters,  Annual Review  Astron. Astrophys. ({\bf 60}, 73 (2022).

\item
A. Siebert, B. Famaey, J. Binney, et al., Mon. Not. R. Astron. Soc. {\bf 425}, 2335 (2012).

\item
M.D.V. Silva, R. Napiwotzki, Mon. Not. R. Astron. Soc. {\bf 431}, 502 (2013).

\item
A. Toomre, {\it The structure and evolution of normal galaxies} (Proc. Advanced Study Institute, Cambridge, England, August 3--15, 1980. Cambridge and New York, Cambridge University Press,  p. 111--136,  1981).

\item
J.P. Vall\'ee, Astrophys. J. {\bf 454}, 119 (1995).

\item
J.P. Vall\'ee, Astrophys. J. {\bf 566}, 261 (2002).

\item
J.P. Vall\'ee, Astron. J. {\bf 135}, 1301 (2008).

\item
J.P. Vall\'ee, International Journal of Astronomy and Astrophysics {\bf 3}, 20 (2013).

\item
J.P. Vall\'ee, Mon. Not. R. Astron. Soc. {\bf 450}, 4277 (2015).

\item
J.P. Vall\'ee, Astrophys Space Sci {\bf 362}, 79 (2017).

\item
Y. Xu, C. J. Hao, D.J. Liu, et al., Astrophys. J. {\bf 947}, 54 (2023).

 \end{enumerate} }
\end{document}